\documentclass[twocolumn,amsmath,amssymb]{revtex4}

\usepackage{graphicx}
\usepackage{dcolumn}
\usepackage{bm}
\usepackage{epsfig,bm,epsf,float,amssymb,amsmath}

\begin{document}

\preprint{UCRL ID}

\title{Monitoring the Thermal Power of Nuclear Reactors with a Prototype Cubic Meter Antineutrino Detector}

\author{A.~Bernstein}
\affiliation{Lawrence Livermore National Laboratory, 7000 East Ave., Livermore,
CA 94550} \email{bernstein3@llnl.gov}

\author{N.~S.~Bowden}
\affiliation{Sandia National Laboratories, 7011 East Ave, Livermore, CA 94550 }

\author{A.~Misner}
\author{T.~Palmer}
\affiliation{Department of Nuclear Engineering, Oregon State University,
Corvallis, OR 97331}

\date{\today}

\begin{abstract}
In this paper, we estimate how quickly and how precisely a reactor's operational status and thermal power can be monitored over hour to month time scales, using the antineutrino rate as measured by a cubic meter scale detector. Our results are obtained from a detector we have deployed and operated at 25 meter standoff from a reactor core. This prototype can detect a prompt reactor shutdown within five hours, and monitor relative thermal power to 3$\%$ within 7 days. Monitoring of short-term power changes in this way may be useful in the context of International Atomic Energy Agency's (IAEA) Reactor Safeguards Regime, or other cooperative monitoring regimes.  

\end{abstract}

\pacs{89.30.Gg 28.41.-i}
\maketitle

\section{Introduction}
\label{sec:intro}

The International Atomic Energy Agency uses an ensemble of procedures and technologies, collectively referred to as the Safeguards Regime, to detect diversion of fissile materials from civil nuclear fuel cycle facilities into
weapons programs. Nuclear reactors are a central element of the nuclear fuel cycle and of the Safeguards Regime. As we show here, it is possible and practical to monitor the operational status and thermal power of reactors with an antineutrino detector. 

In the context of cooperative monitoring, an independent measure of the reactor power can allow confirmation of normal operation of the reactor without a physical inspection, and places a constraint on the total amount of fissile material generated in a given period. The measurement can also be used to verify the operator's own declarations of the reactor power and fuel burnup. 

In an earlier paper \cite{firstpaper}, we presented a general method for exploiting the high rate of antineutrinos emitted by fission reactors to track the power and plutonium content of the reactor core in real time. Such monitoring was first performed by a Russian group at a reactor in Ukraine \cite{Klimov}. Recently, we presented first results from a detector, ``SONGS1'', developed to demonstrate this method as a possible non-intrusive, remotely operated safeguards tool \cite{secondpaper}. SONGS1 has been acquiring data at the San Onofre Nuclear Generating Station (SONGS) in Southern California over the past two years. With it, we have been able to confirm many of the important claims made in \cite{firstpaper}: non-intrusiveness with regard to core and site operations; continuous, remote and automatic data collection and calibration; sensitivity to both short term (several hour) and medium term (daily or weekly) reactor power excursions; and sensitivity to changes in the reactor fissile isotopic content. 

In this paper, we quantify sensitivity of our prototype to relative changes in the reactor power over hourly to monthly time scales, using only the antineutrino signal. Over these time scales, we show this sensitivity is limited primarily by counting statistics even with a detector of quite simple design.

The relevance of the measurement for reactor safeguards depends on the period over which the antineutrino data is acquired. One can monitor the relative power on an hourly basis and look for sudden outages or other short-term anomalies in reactor operations. This type of monitoring may be of interest for off-line refueled reactors, since a reactor outage allows the operator direct access to the weaponizable fissile material in the core. Alternatively, one can acquire data for week or month long periods, and measure the average power in this period with the higher precision afforded by the longer integration time. This more precise measurement can be used to verify stable operations. For continuous monitoring throughout a typical 12-24 month reactor cycle, the month-to-month antineutrino rates must be corrected to account for the influence of variations in the isotopic content of the reactor fuel. This effect is analyzed in a forthcoming companion paper.

It is important to emphasize that antineutrino-based monitoring need not depend on operator declarations - the  detector can be kept under the control of the safeguards agency, providing a wholly independent measurement of reactor status. The non-intrusive and continuous nature of the antineutrino signal, the fact that it provides quantitative information about the reactor thermal power and burnup, is under control of the safeguards agency, and does not require frequent site visits, all point to its potential utility for cooperative monitoring.

\section{Antineutrino Emission from Nuclear Reactors}
\label{sec:production}

Antineutrino emission in nuclear reactors arises from the beta decay of neutron-rich fragments produced by heavy element fissions, and is thereby linked to the fissile isotope production and consumption processes of interest for reactor safeguards. On average, a fission is followed by the production of approximately six antineutrinos. The antineutrinos emerge from the core isotropically, and effectively without attenuation. Over the few MeV energy range within which reactor antineutrinos are typically detected, the average number of antineutrinos produced per fission is significantly different for the two major fissile elements, $^{235}\textrm{U}$ and
$^{239}\textrm{Pu}$. Hence, as the core evolves and the relative mass fractions and fission rates of these two elements change, the measured antineutrino flux in this energy range will also change. This relation between the fissile mass fractions and antineutrino flux, known as the burnup effect, has been observed consistently in previous experiments, e.g. \cite{KamLAND}. 

For our present purpose it is useful to express the relation between fuel isotopics and the antineutrino count rate
explicitly in terms of the reactor thermal power, $P_{th}$. The thermal power is defined as
\begin{equation}
P_{th} = \sum_{i} N^{f}_{i} \cdot E^{f}_{i},
\label{eq:power}
\end{equation}
where $N^{f}_{i} $ is the number of fissions per unit time
for isotope $i$, and $E^{f}_{i}$ is the thermal energy released per fission for this
isotope. The sum runs over all fissioning isotopes, with $^{235}\textrm{U}$, $^{238}\textrm{U}$,
$^{239}\textrm{Pu}$, and  $^{241}\textrm{Pu}$ accounting for more than $99\%$ of all fissions. 

For reactor power monitoring applications, it is important to note that the thermal energy release per fission $E^{f}_{i}$ differs from the \textit{total} energy release per fission, which includes contributions $E_{\nu}$ from the antineutrinos themselves, from neutron capture on fission products $E_{nc}$, and a time dependent term $\Delta E_{\beta \gamma}$ arising from beta and gamma decays which have not completed by a given instant in time. Fortunately, the neutron capture component is readily calculable from instantaneous fission product inventories, and the relative contribution from decaying betas and gammas is small. As a result, the thermal power is nearly proportional to the fission rate as defined in equation \ref{eq:power}. \cite{Kopeikin} calculates the ratio of these terms to be $E^{f}_{i}:E_{\nu} :\Delta E_{\beta \gamma}:E_{nc} \simeq 200:9:0.3:10$ 

Following the formulation in \cite{Huber}, we define the power fractions $f_{i}(t)$ contributed by each isotope as

\begin{equation}
f_{i}(t) =\frac{N^{f}_{i}(t)  \cdot E^{f}_{i}}{P_{th}}.
\label{eq:powerfracs}
\end{equation}

The antineutrino emission rate $n_{\bar{\nu}}(t)$ can then be expressed in terms of the power fractions and the total thermal power as:

\begin{equation}
n_{\bar{\nu}}(t)  = P_{th}(t) \sum_{i} {\frac{f_{i}(t)}{E^{f}_{i}}} \int
dE_{\bar{\nu}} \phi_{i}(E_{\bar{\nu}}) , 
\label{eq:nu_prod_rate}
\end{equation}
where the explicit time dependence of the fission fractions and, possibly, the
thermal power are noted. $\phi(E_{\bar{\nu}})$, is the energy dependent antineutrino
number density per MeV and fission for the $i$th isotope. $\phi(E_{\bar{\nu}})$ has been measured and tabulated by various authors, with recent summary tables available in \cite{Huber}. 

Equation \ref{eq:nu_prod_rate} defines the burnup effect. The fission rates $N^{f}_{i}(t) $ and  power fractions $f_{i}(t)$ change by several tens of percent throughout a typical reactor cycle as $^{235}\textrm{U}$ is consumed and $^{239}\textrm{Pu}$ produced and consumed in the core. These changes directly affect the antineutrino emission rate $n_{\bar{\nu}}(t)$.

\section{Antineutrino Detection Through Inverse Beta Decay Interactions} 

Reactor antineutrinos are normally detected via the inverse beta decay process on quasi-free protons in hydrogenous scintillator. In this charged current interaction, the antineutrino $\bar{\nu}$ converts the proton into a neutron and a positron: $\bar{\nu} + p \rightarrow e^{+} + n$. For this process, the cross section $\sigma$ is small, with a numerical value of only $\sim10^{-43} cm^{2}$. The small cross section can be compensated for with an intense source such as a nuclear reactor. For example, cubic meter scale hydrogenous scintillator detectors, containing $\sim 10^{28}$ target protons $N_{p}$, will register thousands of interactions per day at standoff distances of 10-50 meters from typical commercial nuclear reactors.

In a measurement time \textit{T}, the number of antineutrinos detected via the inverse beta decay process is: 

\begin{equation}
N_{\bar{\nu}}(t)  = \left(\frac{T N_{p}}{4 \pi D^{2}}\right) P_{th}(t) \sum_{i}
{\frac{f_{i}(t)}{E^{f}_{i}}} \int dE_{\bar{\nu}} \sigma \phi_{i} \epsilon.
\label{eq:nu_det_rate}
\end{equation}

In the above equation, $\sigma$ is the energy-dependent cross-section for the inverse beta decay interaction, $N_{p}$ is the number of target protons in the active volume of the detector, and $D$ is the distance
from the detector to the center of the reactor core. $\epsilon$ is the intrinsic detection efficiency, which may depend on both energy and time. The antineutrino energy density and the detection efficiency are folded with the cross-section $\sigma$, integrated over all antineutrino energies, and summed over all isotopes $i$ to yield the antineutrino detection rate. 

To further clarify the relation between the thermal power, the fuel burnup, and the antineutrino detection rate, it is useful rewrite this equation as:

\begin{equation}
N_{\bar{\nu}}(t)  = \gamma \left(1+k(t)\right) P_{th}(t), 
\label{eq:nu_det_rate2}
\end{equation}
where $\gamma$ is a constant encompassing all non-varying terms, including the number of target protons, the detector standoff distance, and the detection efficiency. $k(t)$ describes the change in the antineutrino
flux due to changes in the reactor fuel composition. $\gamma$ is chosen so
that the value of $k$ at the beginning of a reactor fuel cycle is zero.

Typically, commercial reactors are operated at constant thermal power. In this
mode, $k$ decreases by $\approx 0.1$ over the course of a reactor fuel cycle,
depending on the initial fuel loading and operating history, i.e. the
antineutrino detection rate decreases by $\approx 10$\%. The magnitude of this
effect can be predicted at the few percent level in an absolute sense,
if the reactor fuel loading and power history are known. Much of the
uncertainty arises from systematic shifts in measured antineutrino energy
densities $\phi_{i}(E_{\bar{\nu}})$, so that the relative uncertainty in the
predicted burnup rate can be considerably smaller. 

\section{Reactor Simulation}

\label{sec:rsim}

In order to study the relation between the reactor thermal power and the measured antineutrino rate, we must quantify the distorting effect of fuel burnup on the antineutrino rate. For this purpose, we simulated the isotopic evolution of the SONGS core through a single fueling cycle of the reactor, using the ORIGEN simulation package \cite{ORIGEN}. ORIGEN benchmarking studies against assayed fuel assemblies have shown that the package predicts the fissile isotopic content of Low Enriched Uranium (LEU) fuel with 1-2$\%$ accuracy \cite{ORIGENprec}.  

\begin{figure}[h]
\centering
\includegraphics*[width=.9\columnwidth]{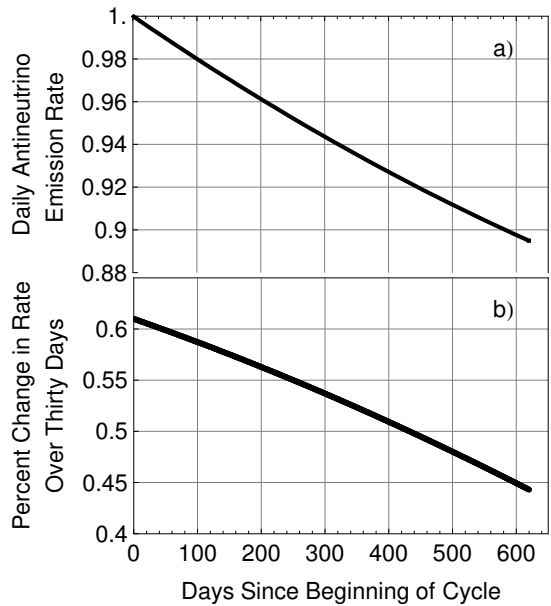}
\caption{ a) The predicted relative daily antineutrino emission rate versus cycle day. The rate is normalized to its value at the beginning of the cycle. b) The percent change in the instantaneous antineutrino emission rate over the previous thirty day period,  versus cycle day.}
\label{fig:nupredict}
\end{figure}

The input fuel loading per fuel assembly, the fuel assembly power densities, and nominal cycle time were obtained from the reactor operator.  The assemblies were simulated in ORIGEN and the fission rates, and mass and number densities of the main fissile isotopes tracked through a 590 full power day evolution, corresponding to the length of Cycle 13 of the SONGS Unit 2 reactor. The fission rates predicted by the simulation, folded with the antineutrino spectral densities \cite{Huber}, allow us to estimate the emitted antineutrino rate at any time during the cycle. (Fig. \ref{fig:nupredict}a). 
 
For the long cycles typical of the San Onofre plant, the burnup effect causes a $10\%$ decrease in detectable antineutrinos by end of cycle. Cycle times have lengthened over the last two decades as plant operations have improved, with the result that the cumulative effect of burnup on the antineutrino count rate is larger. However, the relevant point for the current analysis is that the effect is small over the hour to month time scales of interest for safeguards-related thermal power monitoring. The effect of burnup is never greater than $0.62\%$ over any thirty day period (Fig. \ref{fig:nupredict}b). Provided that the detection efficiency, target mass and distance are constant, this figure demonstrates that the change in the quantity $k(t)$ in Equation \ref{eq:nu_det_rate2} is small over month or shorter time scales, so that the detected antineutrino rate is nearly proportional to reactor power over these time intervals.

\section{The SONGS Reactor}
\label{reactorsection}

The SONGS1 detector is deployed at Unit 2 of SONGS.
There are two operational reactors at this station;
both are pressurized water reactors designed by
Combustion Engineering in the 1970s and have
maximum thermal (electric) power of 3.4 GWt
(1.1 GWe). The detector is located in the tendon gallery of
Unit 2. A feature of many commercial reactors, the
tendon gallery is an annular concrete hall that lies
beneath the walls of the reactor containment structure.
It is used to inspect and adjust the tension in
reinforcing steel cables known as tendons, which extend throughout the
concrete of the containment structure. At SONGS
these inspections occur every several years, and involve
the examination of only a handful of representative
tendons. Apart from these inspections, the
placement of the detector in the tendon gallery has
little or no impact on regular plant operations, and
vice-versa.

SONGS1 is located 24.5 $\pm $ 1.0 m from the Unit 2
reactor core and 149 $\pm $ 3 m from that of Unit 3. The physical core is well approximated by a 4 m tall cylinder with a 3.5 m diameter. Compared with a point source, its finite size has a less than 1$\%$ effect on the measured rate at the 24.5 m standoff distance of the SONGS1 detector.

Since the antineutrino flux generated by each core is
isotropic, 97$\%$ of the reactor antineutrinos reaching
the SONGS1 detector originate from Unit 2. With
Unit 2 at full power the antineutrino flux at the
SONGS1 location is $\displaystyle~10^{17} m^{-2}s^{-1}$.

\section{The SONGS1 Detector}

As described in detail elsewhere \cite{secondpaper}, the SONGS1 detector consists of three subsystems; a central detector, a
passive shield, and a muon veto system. 
The central detector consists of four identical stainless steel cells filled with a total
of 0.64 $\pm$ 0.06 tons of liquid scintillator.  A passive water/polyethylene shield for gamma rays and neutrons surrounds the detector on six sides, with an average thickness of 0.5 m. A 2 cm thick plastic scintillator envelope read out by PMTs covers five sides of the detector and identifies cosmic ray muons.

In the central detector, positrons created by the inverse beta process deposit energy via Bethe-Bloch ionization
as they slow in the scintillator. Annihilation with an electron yields two gamma rays
which can deposit up to an additional 1.022 MeV of energy in the detector. This set of interactions, occuring within about 1 ns, is referred to as the ``prompt" energy deposition. The neutron carries away a few keV of energy from
the antineutrino interaction. After thermalization, this neutron can be detected by capture on a gadolinium (Gd) dopant. A concentration
of  0.1$\%$  Gd by weight yields a neutron capture
time of 28 $\mu$s, and an 8 MeV energy release via
a gamma ray cascade resulting from the capture. The measured response of the detector to this 8 MeV cascade is referred to as the ``delayed" energy deposition. 

The time separation between prompt and delayed energy depositions follows an exponential distribution
with time constant equal to the 28 $\mu$s neutron capture time in the Gd doped scintillator. Taken together, the prompt and delayed energy depositions are referred to as a \textit{correlated} event. This refers to the fact that the 
same underlying physical process generates both interactions, and that they occur close in time relative to most other pairs of interactions taking place in the detector. 

Correlated events can be created by mechanisms other than inverse beta decay. For example, fast muogenic neutrons can scatter off protons in the scintillator, giving a prompt energy deposition, and then be captured on Gd with the same time distribution as occurs for inverse beta events. Such events, which mimic the time structure of the antineutrino events, are referred to as correlated backgrounds. The detector also registers uncorrelated backgrounds -  random coincidences
between two energy depositions from natural radionuclide decays and other sources. Since the occurrence
of such uncorrelated backgrounds is governed by Poisson statistics, the time separation between
these events will also follow an exponential distribution, with a time constant equal to the inverse of the
single event rate (effectively the detector trigger rate).

The detector trigger rate above a 1 MeV threshold, and the trigger rate for the muon veto system are both $\sim$ 500 Hz.
Data is acquired through a NIM/VME-based Data Acquisition System and an on-site computer automatically performs the data analysis. A telephone modem is used to automatically retrieve the results of this analysis, as well as detector state of health indicators, allowing for the remote monitoring of detector operation, and, through the antineutrino signal, of reactor operation. The prototype SONGS1 system (which is not a highly engineered design) operates unattended for months at a time.

\subsection{Expected Antineutrino Interaction Rate}

With our current 0.64 ton liquid scintillator detector at a standoff of 24.5 $\pm$  1 m from the reactor
core, and with the reactor parameters defined in Sec. \ref{reactorsection}, the rate predicted by Eqn. \ref{eq:nu_det_rate2} at the beginning of the reactor fuel cycle ($k(t) = 0$) is  3800 $\pm$  440
antineutrino interactions per day for $100\%$ detection efficiency. The uncertainty in the absolute antineutrino rate arises primarily from the uncertainty in our knowledge of the amount of scintillator in the detector. This uncertainty is a systematic shift in the absolute rate that has no effect on the relative power measurements of interest in the analysis presented below.  

\section{The Selection Procedure for Antineutrino Events}

To isolate antineutrino events, we form candidate event pairs from the raw data. For each pair of sequential
events meeting the hardware trigger criteria, the first is labeled prompt or positron-like,
and the second  delayed or neutron-like. Two
time intervals are also defined: the interval between 
the prompt event and the most recent muon trigger or central detector 
trigger, and the time between the
prompt and delayed events (interevent time).
To select antineutrino candidates we apply several
cuts to the sequential pairs. To exclude events with highly non-uniform light collection, a cut is applied to  ensure that each PMT observing a particular cell sees a similar amount of light per event. For two PMTs observing a given cell, a and b, the cut is applied to the ratio z, defined as:

\begin{equation}
z =  \frac{a- b}{a+ b} < 0.4. 
\end{equation}

This selection criterion is referred to as the
ratio cut. In practice, it determines which interaction locations in the scintillator are accepted by the selection process, thereby defining a fiducial volume. 

Next, we apply cuts on the amount of energy
recorded in each cell. For simplicity and flexibility
in our analysis, we have chosen to consider
each cell as an independent detector.

Hardware limitations impose a cut of 10 $\mu $s on the minimum time between any event pair. Finally, we accept only those events that occur at least $100~{\mu}$s after the last muon hit/acquisition trigger. This final cut significantly reduces the number of contaminating antineutrino-like event pairs in the data set caused by muon interactions. From the empirically measured time constant of the muon correlated events (20 $\mu$s), we calculate that this cut excludes all but 0.7$\%$ of antineutrino-like backgrounds correlated with a muon recorded in the veto.

Table~\ref{tab:cuttab} summarizes our  event selection criteria. The predicted rate of antineutrinos at beginning of cycle based on these criteria is $407\pm75$~/day. As mentioned earlier, this error includes the large absolute uncertainty in the number of target atoms. As shown below, a relative measurement has a considerably smaller uncertainty. 

\begin{table}
\caption{The Antineutrino Event Selection Criteria} \label{tab:cuttab}
\begin{tabular*}{1.0\columnwidth}{l r r} \hline
Quantity       				&  Criterion  			     \\ 
\hline
Prompt Energy ~	&~$2.39~MeV < E_{positron} < 9~MeV$ 	    \\
Delayed Energy~	&~$3.5~MeV < E_{neutron} < 10~MeV$ 	    \\
PMT Ratio Cut   		&~~$|z| <~0.4$				\\ 
Interevent Time 	&~$t_{min}=10~\mu$s			\\
Muon Veto		&~$t_{mu}>100~\mu$s			\\
\hline
\end{tabular*}
\end{table}

\section{The Antineutrino-Like Event Sample}

\begin{figure}[h]
\centering
\includegraphics*[width=\columnwidth]{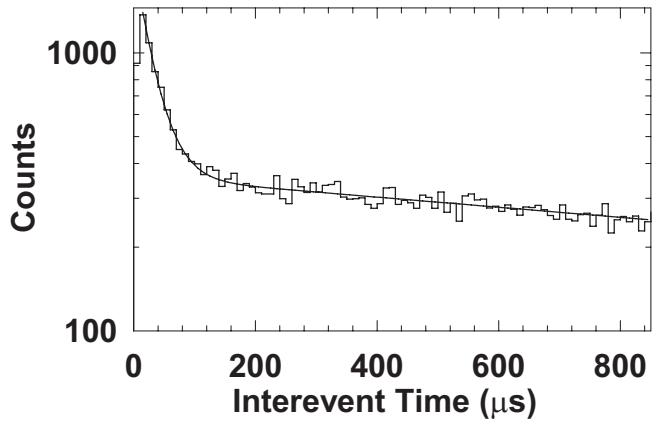}
\caption{ A representative interevent time spectrum of event
pairs that pass all selection cuts acquired during 7 days of
data taking. The fit to the data has four parameters: an exponent and amplitude each for the fast and slow exponentials describing respectively the correlated signal and coincident singles background.}
\label{fig:nutime}
\end{figure}

After applying these selection criteria, we examine
the spectrum of time intervals between pairs of successive energy depositions (Fig. \ref{fig:nutime}). Two clear exponential features are visible. The faster and more prominent of these arises from a prompt, positron-like, energy deposition followed a characteristic time later by a delayed, neutron capture-like energy deposition. The second, slower exponential is due to the random coincidence of two sequential background events both of which exceed the relevant thresholds. As would be expected for such coincidences, the time constant of this slow exponential is equal to the inverse of the acquisition trigger rate.

To further select antineutrino events we integrate the interevent time distribution from $10~{\mu}$s to $100~{\mu}$s. Beyond the upper time limit of $100~{\mu}$s, the event sample is dominated by uncorrelated backgrounds. This integral defines the total number of events prior to any background subtraction, and includes contributions from true antineutrino interactions as well as both correlated and uncorrelated backgrounds. We refer to these as antineutrino-like events. 

It is impossible in this detector to distinguish backgrounds from true antineutrino interactions on an event-by-event basis. For the analysis below, when a net number of antineutrino events is required, we perform a statistical subtraction of the same definite integral determined over a reactor off period. We use this procedure to estimate the net daily or weekly number of antineutrinos. For integration times less than 24 hours, unacceptably large uncertainties are introduced by such a procedure due to limited counting statistics. Instead, we apply a standard hypothesis test to the total (non-background-subtracted) antineutrino-like event rate to look for signifcant changes.

\section{Hourly Monitoring of the Reactor Operational Status}
\label{sec:opstatus}

In the off-line refueled reactors which constitute the majority of the world's reactor stock under safeguards, fissile materials can only be accessed during shutdowns. Thus, it is of potential interest for safeguards to estimate how quickly a shutdown or a large change in the reactor thermal power can be identified using the antineutrino signal.

For a quantitative estimate of the statistical significance of such changes, the Sequential Probability Ratio Test (SPRT) \cite{Wald} can be used to estimate the amount of time required to conclude, with a given level of confidence, that the reactor thermal power has changed based on a change in the antineutrino signal. The SPRT is based on a log-likelihood ratio, defined as 

\begin{equation}
r_{SPRT}=log{\frac{P(\mu_{1},nevents)}{P(\mu_{2},nevents)}}.
\label{eq:sprt}
\end{equation}
$P(\mu,nevents)$ is the probability that the measured number of events $nevents$ is drawn from a parent distribution with mean $\mu$. $\mu_{1}$ and $\mu_{2}$ are the expected mean values of the number of events in a fixed time interval during periods at two different power levels - for example, 100$\%$ power and zero power. In the examples discussed below, the expected mean values are derived from data taken during periods in which the thermal power was known by independent means to be at the level being tested for (zero power, full power, or some intermediate value). In a real regime, a set of calibrations between thermal power and antineutrino rate would be established in an initial cycle, and used at later times to specify mean values $\mu$ which would serve as inputs for the test statistic. Using the test, measured mean values at any given time would be tested against a set of possible expected values derived from the earlier calibration. The length of the calibration periods would be optimized based on the detected antineutrino event rate, and the size of the burnup effect for a given reactor and fuel type. With the configuration discussed here, we used a calibration period of one month, which gives  $\sim 1\%$ statistical accuracy on the rate, compared with a $<0.65\%$ systematic error induced by burnup, as described earlier. 

To apply the test, one calculates the cumulative sum of the logarithmic quantity $r_{SPRT}$, with the sum updated at fixed intervals. Once this sum exceeds an upper or lower bound $a$ and $b$, the test has confirmed that the reactor is in one and only one of the two possible states, with a specified level of confidence. The statistic is then reset and the testing process continues anew until a conclusion is reached. This is referred to as `online' testing, since the test is constantly being updated with new data until a decision is made, at which point the statistic is reset.

The probability for change detection can be directly quantified in terms of a change in the test statistic. The upper and lower thresholds are defined in terms of the probability $\alpha$ of a false alarm, and probability $\beta$ of a false negative (i.e. failing to recognize the change of state when it has occurred). The equations relating the thresholds for the test statistic to the probabilities of detection or non-detection are:

$$ a = \log{\frac{1-\beta}{\alpha}} \mbox{and }  b= \log{\frac{1-\alpha}{\beta}}.  $$
The thresholds can be selected to balance the need for timely detection with the need for high confidence in the result.  For the analysis below, we demand 1$\%$ probability of either a false alarm or false negative.

Under a broad range of conditions, this method allows one to detect a change in the status of a time-varying process within the minimum possible time \cite{Wald}. With the SPRT method, one can also explicitly quantify the probability of false alarms (changes being detected where none existed) and misses (undetected changes in operational status) \cite{Smith}.

\begin{figure}[h]
\centering
\includegraphics*[width=0.9\columnwidth]{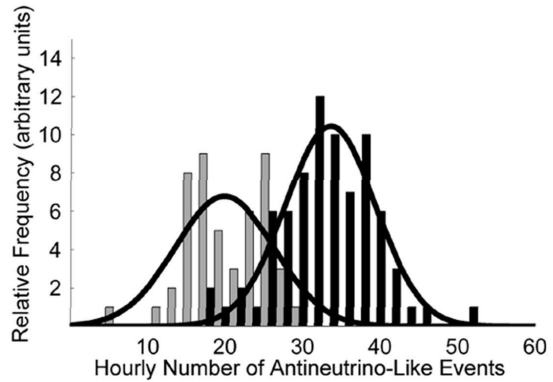}
\caption{ A histogram of the number of antineutrino-like events per hour interval before (black) and during (light gray) the reactor outage.}
\label{fig:onoffpoiss}
\end{figure}

A simple implementation of the test assumes Poisson distributed signal and background. We first confirm that our data meet this criterion. Fig. \ref{fig:onoffpoiss} shows the histograms of the number of detected antineutrino-like events in one hour intervals for equal amounts of reactor on and reactor off data. Both distributions are reasonably well fit by a Gaussian distribution, a good approximation for the Poisson distribution even with these low counting statistics. The chi-squared per degree of freedom values for the Gaussian fits are 0.9 and 1.1 for the on and off data respectively. As seen in the figure, there is significant overlap in the distributions, so that a single measurement is insufficient to determine the operational status. 

\begin{figure}[h]
\includegraphics*[width=1\columnwidth,viewport=60 72 500 600]{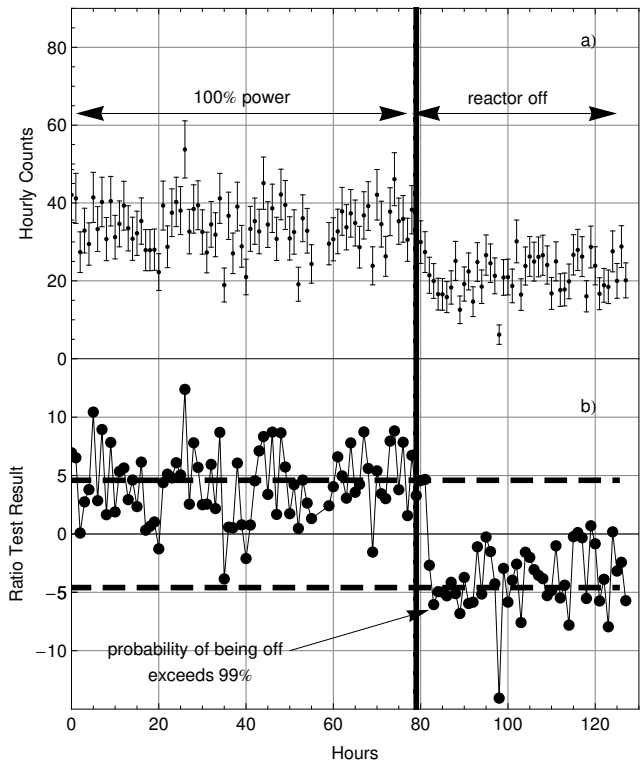}
\caption{ (a): the hourly number of antineutrino-like events, plotted versus hour, through a reactor outage. (b): the value of the SPRT statistic plotted versus hour over the same time range. The dashed lines in this figure are the 99 $\%$ confidence level values of the test statistic. In both plots, the vertical line indicates the hour in which the reactor shutdown occurred. Values of the statistic above the upper dashed indicate that the reactor is in the on state. For this data set, these values are obtained only during the reactor on period, meaning that no false positives or negatives occurred.  }
\label{fig:scramctssprt}
\end{figure}

In our first example, the outage occurs within one hour (Fig. \ref{fig:scramctssprt}a), and the test requires about 5 hours of data to determine that the reactor has turned off, with 1$\%$ probability of either a false alarm or false negative (missed alarm) (Fig. \ref{fig:scramctssprt}b). This condition is satisfied when the value of the test statistic is smaller than $b=-4.59$. 

\begin{figure}[h]
\centering
\includegraphics*[width=1\columnwidth,viewport=60 72 575 650]{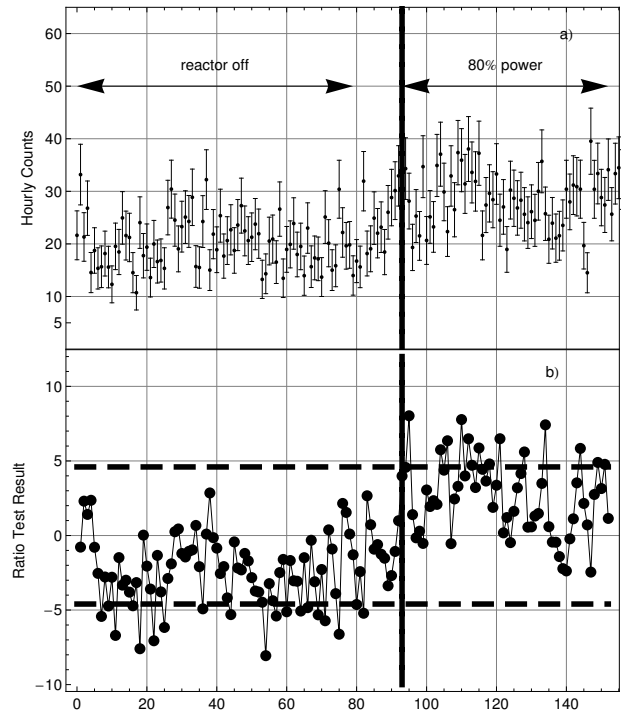}
\caption{(a): the hourly number of antineutrino-like events, plotted versus hour, through a reactor ramp from  zero to 80$\%$ power. (b): the value of the SPRT statistic plotted versus hour over the same time range. The dashed lines have significance analogous to those in Fig. \ref{fig:scramctssprt}.}
\label{fig:rampto80sprt}
\end{figure}

In another excursion, the reactor was ramped from zero to 80$\%$ power over an $\sim$14 hour period, and held at this level for 3 days (Fig. \ref{fig:rampto80sprt}a). In this case, the test requires 2 hours to determine that the reactor is in fact in the 80$\%$ power state, with 99$\%$ confidence. (Fig. \ref{fig:rampto80sprt}b). 

\begin{figure}[h]
\centering
\includegraphics*[width=\columnwidth,viewport=45 45 575 625]{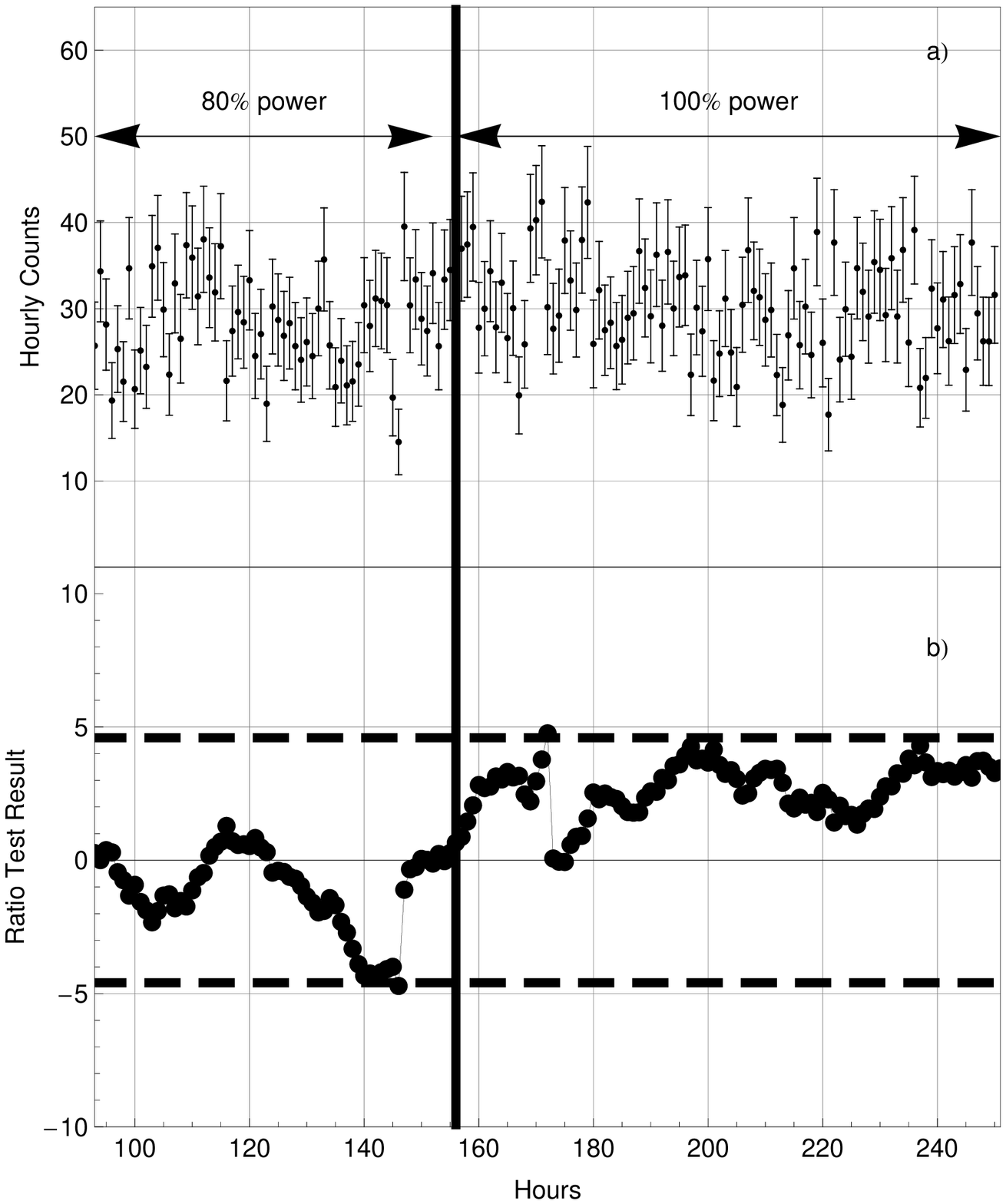}
\caption{(a): the hourly number of antineutrino-like events, plotted versus hour, through a reactor ramp from  80$\%$ to 100$\%$ power. (b): the value of the SPRT statistic plotted versus hour over the same time range. The dashed horizontal lines have significance analogous to those in the lower plot in Fig. \ref{fig:scramctssprt}. The solid vertical line shows the hour in which $100\%$ thermal power was obtained.}
\label{fig:rampto100sprt}
\end{figure}

A somewhat more challenging circumstance is presented by a subsequent transition from 80$\%$ to 100$\%$ thermal power. In this case, it is difficult to see the power step by direct inspection of the change in the number of antineutrino-like events (Fig. \ref{fig:rampto100sprt}a). However, the SPRT is able to detect the change with 99$\%$ confidence, albeit with a longer time to detect due to the closer proximity of the mean number of antineutrino-like events in the 80$\%$ and 100$\%$ power states. The test is able to detect the 20$\%$ power shift in approximately twelve hours.(Fig. \ref{fig:rampto100sprt}b).

\section{Daily and Weekly Monitoring of Relative Thermal Power}
\label{sec:thermalpower}

Hourly monitoring can quickly detect gross changes in the operational state of the reactor, such as the 20$\%$ power shift just described. For more precise measurements, longer integration times are required. We define a relative thermal power estimator by forming the ratio of the daily or weekly average antineutrino detection rate (after background subtraction)to the prior month background-substracted average rate. This ratio is proportional to the daily or weekly average of the thermal power, with an accuracy to be calculated below. Assuming the thermal power of the reactor to be known by independent means at the start of a month-long measurement period, the antineutrino rate can thereby be directly associated with a particular power level, and excursions from this level can be detected by changes in the daily or weekly average

Month-long periods of comparison form a natural break point for the analysis presented here. This is due to the fact that the effect of burnup on the detected (and emitted) antineutrino rate never exceeds 0.65$\%$ over one month, as demonstrated in section \ref{sec:rsim}, while statistical uncertainty is at the 1$\%$ level for this averaging period. As long as the period of comparison is no longer than one month, we will show that the variation in the ratio just defined will be dominated by counting statistics, and not by burnup-induced changes or other systematic effects. For longer comparison periods, the effect of burnup on the relative power estimate grows, and we must explicitly introduce our burnup model. This analysis is performed separately in a forthcoming paper.  

The precision of the daily or weekly estimator is determined by counting statistics, and by any time-dependent systematic effects (other than burnup) which can alter the detection efficiency. By making the measurement relative to an initial monthly average value, time independent corrections to the detection rate - such as those caused by an incorrect estimate of the number of targets or of the overall detection efficiency - only induce an overall shift in the constant of proportionality between reactor power and the absolute antineutrino detection rate. This shift has no effect on the stability or precision of our relative power estimator.

To analyze the correlation between the daily and weekly average antineutrino detection rate and the thermal power, we must subtract the background, which is measured during reactor off periods.   

\subsection{Reactor Off Data}
\begin{figure}[t]
\centering
\includegraphics*[width=\columnwidth]{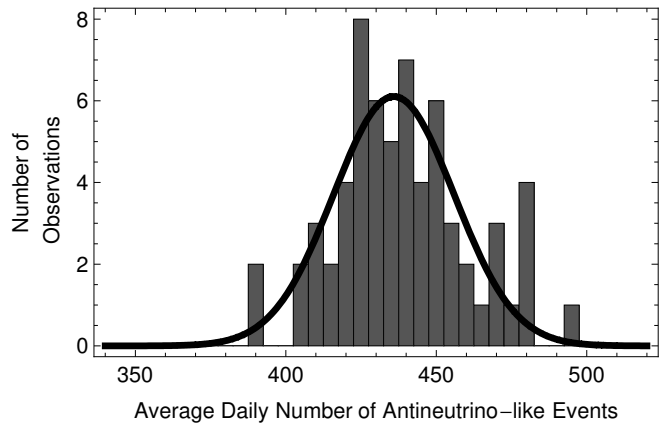}
\caption{A histogram of events passing all antineutrino selection cuts during a 63 day long reactor off period.} \label{fig:offdata}
\end{figure}

During a 63 day period with Unit 2 (the near reactor) at $0\%$ power, the average daily rate of events passing all antineutrino selection criteria was 441 (Fig.~\ref{fig:offdata}). This rate is primarily comprised of uncorrelated non-antineutrino backgrounds (as well as an $\sim$10 count per day contribution of real antineutrinos from the distant reactor). The measured standard deviation for the reactor off data sample is 22.3 events, or 5$\%$, (with an uncertainty in this standard deviation value of 0.5$\%$). The predicted uncertainty due to Poisson counting statistics alone is 4.7$\%$, consistent with the measured standard deviation value. We also performed a Gaussian fit to the data as a check on the measured sample mean and standard deviation. The value of the Gaussian mean is 435 events, close to the sample mean, and the Gaussian sigma is 20 events, close to the sample standard deviation value of 22.3 events.

Over the same 63 day (reactor off) interval, the average weekly rate of events passing all antineutrino selection criteria was 3088, with a measured standard deviation of 88 events, or $2.6\pm0.5\%$. The predicted Poisson uncertainty based on counting statistics in this case is $1.9\%$, close to the sample standard deviation, but indicative of a possible small additional uncertainty not accounted for by counting statistics. The Gaussian mean (3088) and standard deviation (87) are also close to the sample mean and standard deviation. 

For both daily and weekly integration times, the mean value of the background is subtracted from the signal during reactor on periods to obtain the net antineutrino detection rate. 
 
\begin{figure}[t]
\centering
\includegraphics*[width=\columnwidth]{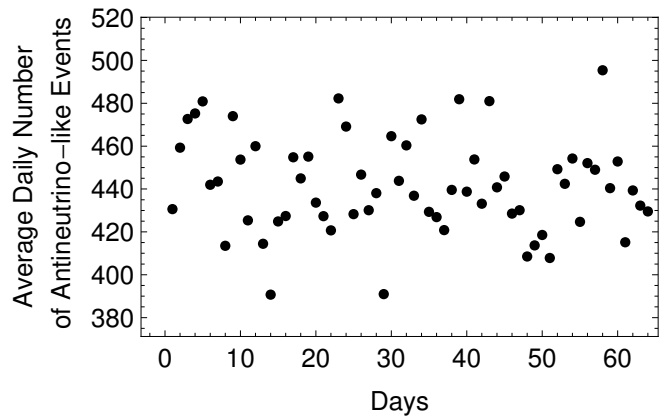}
\caption{Events per day passing all antineutrino selection cuts during a 63 day long reactor off period, plotted versus day.} 
\label{fig:offvtime}
\end{figure}

 A linear fit to reactor off daily antineutrino-like detection rate plotted versus day (Figure \ref{fig:offvtime}) has a slope of $0.13 \pm 0.15$, consistent with a constant background.

\subsection{The Stability of the Antineutrino-Based Power Estimate}

Next we consider the stability of a relative power measurements based on the daily and weekly average background-subtracted antineutrino detection rate. According to operator records, the reactor power in our analysis period is constant to within $0.5\%$, at $99.5\%$ of full power.

Rearranging Equation \ref{eq:nu_prod_rate} it can be seen that the thermal power depends on the measured number of antineutrino events per unit of time, divided by the detector related constant $\gamma$:

\begin{equation} 
P_{th} =\frac{N_{\bar{\nu}}}{\gamma (1+k(t)) }
\label{eq:psimple}
\end{equation}

We define a relative power measurement r as 

\begin{equation}
 r = \frac{<P_{th}>_{day,week}}{<P_{th}>_{month}}  
\end{equation}

The brackets and subscript indicate averaging for the previous day, week or month. The small systematic variation between the daily or weekly value of $k$ and its monthly average can be neglected, and the term $\gamma$, depending on the detector mass and distance, is assumed constant. With better than $1\%$ precision, the variation in the ratio now depends only on the statistical variations in the count rates, averaged over the periods in question. However, this neglects possible time variations in the detection efficiencies over the relevant time interval. If these variations are large, the measured standard deviation of the data will be larger than the $\sqrt{nevents}$ spread expected from Poisson counting statistics alone. We consider this possibility directly.   

With the above assumptions, the ratio r can be expressed in terms of the ratio of average detected antineutrino rates as:

\begin{equation} 
r =\frac{<N_{\bar{\nu}}>_{day,week}}{<N_{\bar{\nu}}>_{month}},
\label{eq:ratio}
\end{equation}
where the subscripted brackets again signify averaging over the indicated time periods. 
 
\begin{figure}[t]
\centering
\includegraphics*[viewport=125 5 375 235]{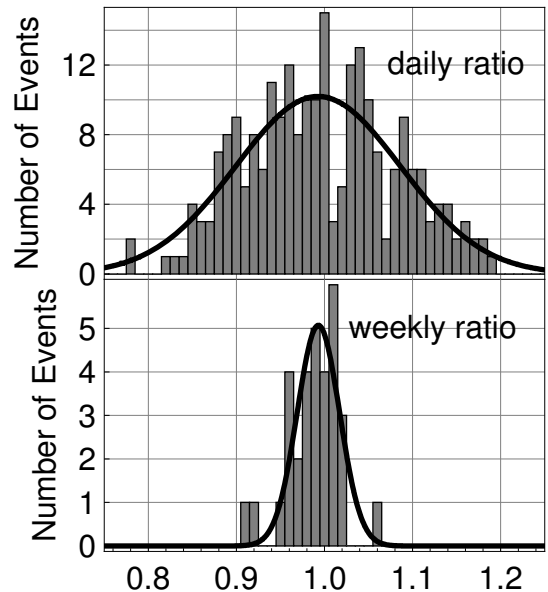}
\caption{The histograms of the ratio of $r$, defined as the net daily or weekly average detected antineutrino rate to the net detected antineutrino rate averaged over the 28 days prior to the measurement. Both the daily and weekly data sets extend over the same 33 week period.} \label{fig:dlywklyratio}
\end{figure}

Fig. \ref{fig:dlywklyratio} shows the spread in this ratio for daily and weekly averaging. By examining the standard deviation of $r$ , we can quantify the degree to which the ratio obeys Poisson statistics, and estimate the contribution of non-Poisson, time-varying systematic drifts on the relative power estimate. 

For the daily averaged data, the measured standard deviation in the ratio $r$ is 8.3$\pm0.5\%$, while the spread based on a gaussian fit is 8.0$\pm0.5\%$. These values are within error of the 7.8$\%$ expected spread in the ratio $r$ due to Poisson statistics alone. For the weekly data,the measured standard deviation in $r$ is 3.0$\pm0.3\%$ - again consistent with the expected 3$\%$ variation due to Poisson statistics. 

For both daily and weekly averaging, statistical uncertainty fully accounts for the total observed spread in the ratio. Aside from the known $\leq0.65\%$ systematic contribution to the spread due to burnup, such small additional effects as are present may come from periodic drifts in the gain scale, or other efficiency changes that are not fully accounted for in the calibration procedure. 

\section{Conclusions}
\label{sec:conclusion}

The above analysis demonstrates that our current prototype detector can monitor changes in reactor status (on versus off) in five hours with greater than 99$\%$ confidence, and can directly measure power levels over month long time scales  with an estimated $8.3\%$ precision using a daily background subtracted number of detected antineutrinos, or $3\%$ using a weekly number, limited almost entirely by statistics. By construction, this estimate  is independent of the long term $\simeq 12\%$ systematic trend induced by the changing core isotopics. It provides a measure of the precision of the detector itself, including all statistical and systematic effects occurring over periods of days to months. Slightly longer integration times, or improvements in the detection efficiency would further reduce this uncertainty. Ultimately, however, the effect of burnup must be  fully accounted to extract a stable long term power or burnup measurement beyond one month. This analysis is the subject of a separate article.  

The fact that our current detector approaches the Poisson limit for a relative power measurement on day to week time scales has further significance. It implies that even a simple detector design can suffice for the relative thermal power monitoring approach envisioned here. Our experience is that the simplicity of the detector design will play a key, even decisive role in determining whether this technology is adopted by the IAEA or other safeguards regimes.

\section*{Acknowledgements}

We thank DOE Office of Nonproliferation Research and Engineering for their sustained support of this project. We are indebted to the management and staff of the San Onofre  Nuclear Generating Station for allowing us to deploy our detector at their site. We thank Norman Madden, Dennis Carr, Alan Salmi, and James Brennan for invaluable technical assistance an advice. We are grateful to Felix Boehm for donation of the liquid scintillator used in our detector, and Giorgio Gratta for his early contributions to this project.

\end{document}